\documentclass[12pt]{article}
\usepackage[utf8]{inputenc}
\usepackage{times}
\title{Spatially entangled Airy photons}
\author{Ohad Lib, Yaron Bromberg$^*$}

\date{}

\usepackage{graphicx}
\usepackage{float} 
\usepackage[sorting = none, backend = bibtex, style=numeric-comp]{biblatex}
\begin{document}

\maketitle
\noindent{Racah Institute of Physics, The Hebrew University of Jerusalem, Jerusalem, 91904 Israel}

\noindent{*Yaron.Bromberg@mail.huji.ac.il}
\begin{abstract}
Over the past decade, Airy beams have been the subject of extensive research, leading to new physical insights and various applications. In this letter, we extend the concept of Airy beams to the quantum domain. We generate entangled photons in a superposition of two-photon Airy states via spontaneous parametric down conversion, pumped by a classical Airy beam. We show that the entangled Airy photons preserve the intriguing properties of classical Airy beams, such as free acceleration and reduced diffraction, while exhibiting non-classical anti-correlations.
\end{abstract}

Airy wave packets, first described by Berry and Balazs in 1979, are free particle solutions to Schrodinger equation that exhibit unique features such as free acceleration and zero diffraction\cite{berry1979nonspreading}. Despite their intriguing properties, Airy wave packets are impossible to realize experimentally since they carry an infinite amount of energy. Almost three decades after their discovery, an approximate solution was suggested in the context of optics, exhibiting quasi-non-diffracting properties\cite{siviloglou2007accelerating}. This solution of the paraxial wave equation, coined Airy beam, shows almost the same remarkable properties as Airy wave packets, such as free acceleration and self healing, while requiring only finite energy\cite{siviloglou2007accelerating,siviloglou2007observation,broky2008self}. Since their first experimental observation over a decade ago using linear optics\cite{siviloglou2007observation}, Airy beams have been demonstrated in various platforms, ranging from nonlinear optics\cite{ellenbogen2009nonlinear,lotti2011stationary} to free electrons\cite{voloch2013generation} and surface plasmons\cite{salandrino2010airy,minovich2011generation}. Following these demonstrations, the unique properties of Airy beams have been utilized in practical applications such as super resolution imaging\cite{jia2014super}, micro-manipulation of particles\cite{baumgartl2008micro}, and the generation of curved plasma channels\cite{polynkin2009plasma}. Recently, single photon Airy beams have been demonstrated as well\cite{maruca2018quantum}.

In this letter, we extend Airy beams to the quantum domain by demonstrating entangled Airy photons. We create highly entangled photons via Spontaneous Parametric Down Conversion (SPDC) by pumping a nonlinear crystal with a classical laser (coined pump beam). By carefully tailoring the shape of the pump beam, we create spatially entangled photons exhibiting Airy shaped quantum correlations, which are not manifested in the distribution of the individual photons. We study the properties of the entangled Airy photons, showing they exhibit the celebrated non-diffracting and free acceleration properties of classical Airy beams, while maintaining their quantum anti-correlated nature.

Classically, a finite energy Airy beam is described by an electrical field envelope of the form $Ai(s)exp(as)$, where $Ai()$ is the Airy function, $s$ is a dimensionless coordinate and $a$ is a small positive parameter ensuring that the beam carries finite energy\cite{siviloglou2007accelerating}. Interestingly, the angular spectrum of the finite energy Airy beam consists of a Gaussian beam modulated by a cubic phase mask, making the experimental realization of such beams relatively simple\cite{siviloglou2007observation}.  

The quantum state of the two entangled photons created in the SPDC process, is given in terms of their transverse wavevector components at the crystal plane $\mathbf{q_s}$ and  $\mathbf{q_i}$ by $\\|{\psi}\rangle=\int d \mathbf{q_s}d\mathbf{q_i}\psi(\mathbf{q_s,q_i})a^\dagger(\mathbf{q_s})a^\dagger(\mathbf{q_i})|0\rangle$. Here, $a^\dagger(\mathbf{q})$ is the creation operator of a photon with a transverse momentum $\mathbf{q}$, $|0\rangle$ is the vacuum state, and the signal and idler photons have the same frequency and polarization. In the thin crystal regime, the two-photon amplitude $\psi(\mathbf{q_s,q_i})$ can be expressed in terms of the angular spectrum of the pump beam $v(\mathbf{q})$\cite{monken1998transfer,walborn2010spatial},
\begin{equation}\label{1}
\psi(\mathbf{q_s,q_i}) = v\left(\mathbf{q_s+q_i}\right).
\end{equation}
Therefore, manipulating the shape of the classical pump beam facilitates control over the two-photon spatial correlations. This interesting correspondence between the entangled photons and the pump beam has been used in various application, such as the generation of maximally entangled orbital angular momentum (OAM) states\cite{torres2003preparation,kovlakov2017bell,kovlakov2018oam}, cancelling the scattering of entangled photons in real time\cite{lib2019generation} and for controlling the spatial coherence of entangled states\cite{defienne2018coherence,giese2018influence}.
Here, we exploit this remarkable fact to generate entangled photons with Airy quantum correlations, by pumping the SPDC process using a classical Gaussian beam with a cubic phase pattern, having angular spectrum of the form $v(q)=Ai(s)exp(as)$, where $s=q/q_0$ is a dimensionless coordinate and $q$ is the transverse component of the pump wavevector. 

\begin{figure}[H]
\centering
\includegraphics[width=0.8\textwidth]{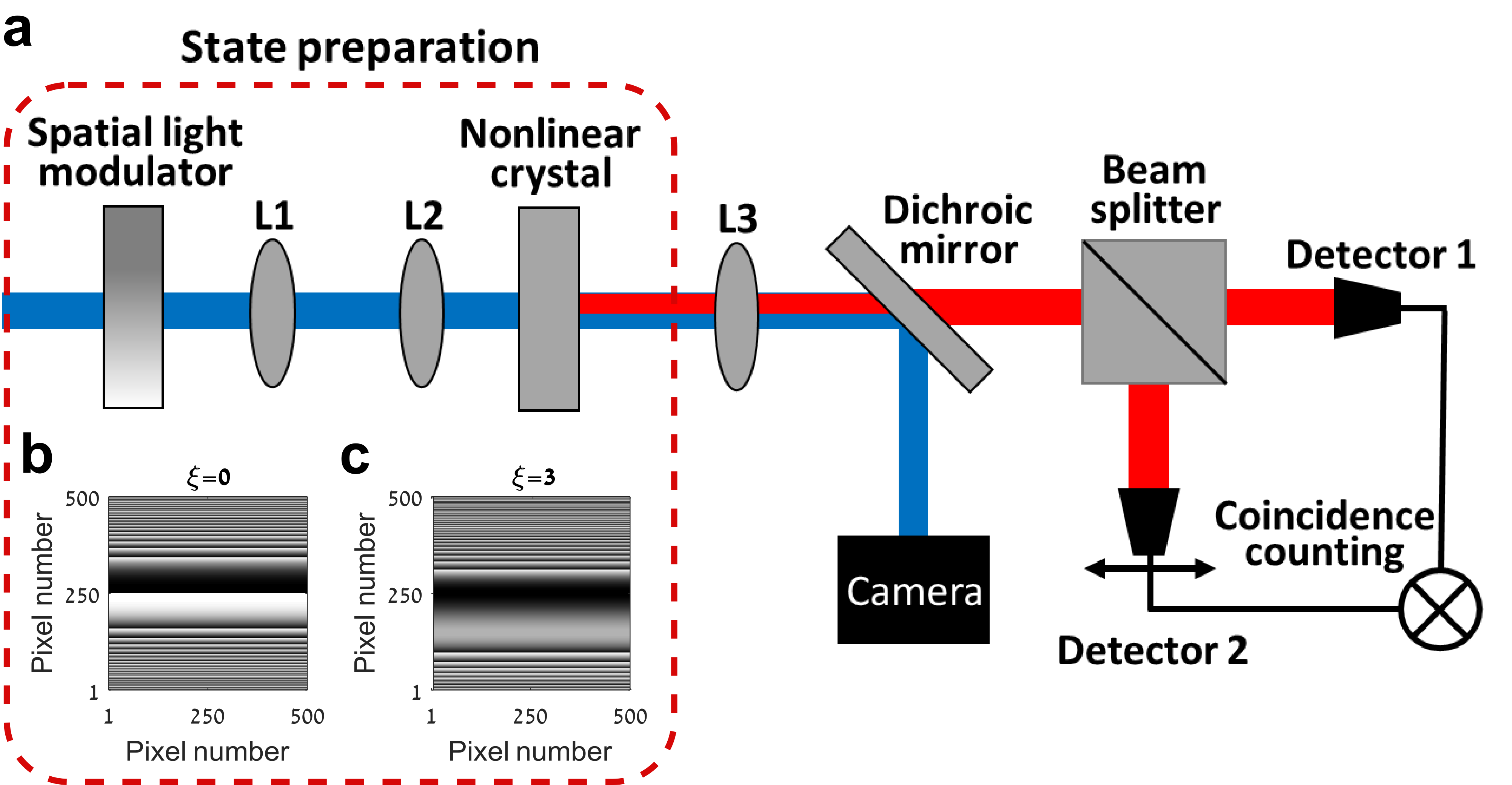}
  \caption{\label{fig:1} Experimental setup. (a) A Gaussian beam emitted by a $404nm$ continues-wave laser is shaped using a cubic phase mask (b) applied by a spatial light modulator (SLM). The shaped classical beam pumps a periodically polled KTP (PPKTP) crystal which is imaged to the SLM plane (using lenses $L1,L2$), creating spatially entangled photons with the desired Airy correlations. Both the classical pump beam and the entangled photons are measured at the far-field after being separated by a dichroic mirror, using a camera and a pair of single photon detectors equipped with $10nm$ bandpass filters, respectively. To study the diffraction of the entangled photons, a controlled quadratic phase is added to the SLM pattern in order to emulate free space propagation (c) (see supplementary). }
\end{figure}

The experimental setup is presented in Fig.\ref{fig:1}. We use a spatial light modulator (SLM) to apply the cubic phase mask onto the Gaussian pump beam (Fig.\ref{fig:1}a,b). The pump beam profile at the SLM plane is imaged onto a $2mm$-thick nonlinear crystal (PPKTP), creating spatially entangled photons via SPDC. 
To measure the quantum correlations between the entangled photons at the far-field, the pump beam is first separated from the entangled pairs using a dichroic mirror, and sent to a camera. The entangled pairs are measured at the far-field of the crystal by using two single photon detectors and a coincidence detection circuit.
Since the angular spectrum of the pump beam, $v(q)$, is given by the Fourier transform of the pump profile at the crystal plane, we get from Eq.\ref{1} that the quantum two-photon amplitude of the entangled photons is given by $\psi(q_s,q_i) = Ai\left(\frac{q_s+q_i}{q_0}\right)exp\left(a\frac{q_s+q_i}{q_0}\right)$.
Therefore, the coincidence pattern at the far-field exhibit an Airy pattern, $C(x_s,x_i)\propto \left| Ai\left(\frac{x_s+x_i}{x_0}\right)exp\left(a\frac{x_s+x_i}{x_0}\right)\right|^2$,where $x_i, x_s, x_0$ are the far-field coordinates corresponding to $q_i, q_s, q_0$, respectively. In our experimental setup, $a=0.1$ and $x_0=175\mu m$.

\begin{figure}[H]
\centering
\includegraphics[width=0.7\textwidth]{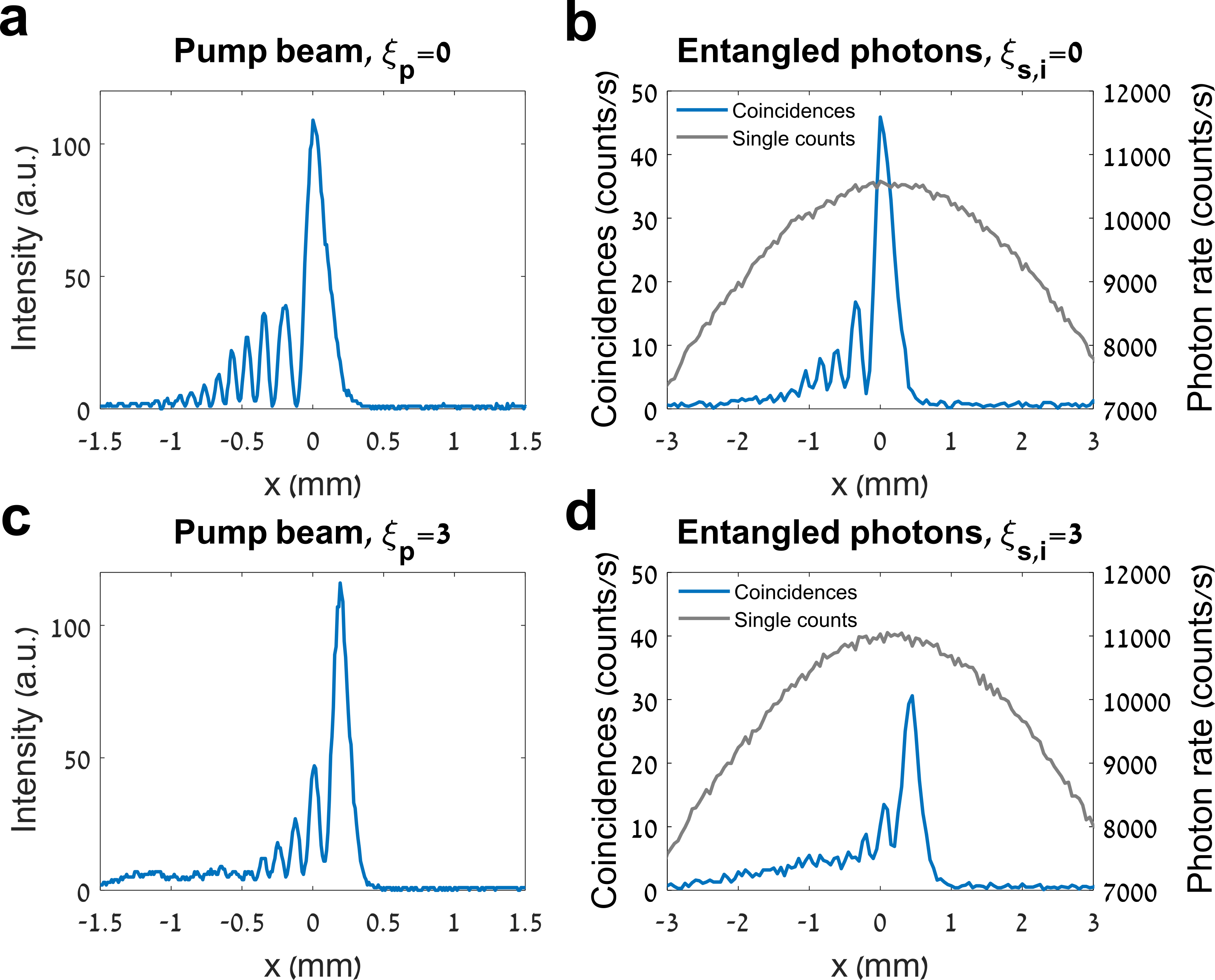}
  \caption{\label{fig:2}  Entangled Airy photons. The pump beam exhibits the desired Airy pattern at the far-field (a). The quantum correlations between the generated entangled photons show an Airy pattern as well, stretched by a factor of two due to the different wavelengths (b, blue curve). The Airy pattern is not manifested in the single photon distribution (b, grey curve). The same measurements are repeated for an emulated propagation distance of $\xi=3$ by adding a quadratic phase function to the SLM. Both the classical pump Airy beam (c) and the Airy entangled photons (d) exhibit negligible diffraction. The Airy pattern is not visible in the single photon distribution also after the propagation.}
\end{figure}

The pump beam far-field intensity profile is presented in Fig.\ref{fig:2}a, exhibiting the desired Airy pattern. As predicted, the measured coincidence pattern of the entangled photons at the far-field follows the same Airy pattern as the pump beam, up to a scaling factor of two, due to the different wavelengths (Fig.\ref{fig:2}b). Interestingly, the single photon distribution is homogeneous and does not exhibit an Airy beam distribution. Only upon detection of one photon, its twin photon collapses to an Airy beam (Fig.\ref{fig:2}b).

One of the remarkable features of classical Airy beams is their accelerating, non-diffracting, propagation.  According to Ehrenfest's theorem, in the absence of any external potentials, the center of mass of an optical beam must be preserved so that the beam propagates in a straight line. However, this restriction does not hold for the local features of the beam, allowing the free transverse acceleration of the local maxima of Airy beams\cite{siviloglou2007accelerating,siviloglou2007observation,greenfield2011accelerating,efremidis2019airy}.  
The intensity profile of a classical Airy beam after propagating a distance $\xi_p=\frac{z}{k_p{x_{0 p}} ^2}$, where $k_p$ is the wavenumber and $z$ is the propagation distance, is given by $\left|Ai(s-(\xi_p/2)^2+ia\xi_p)exp(as-a\xi_p^2/2)\right|^2$\cite{siviloglou2007accelerating,siviloglou2007observation}. For small $a$, the local features of the Airy beam are not significantly spread even over long propagation distances, and accelerate according to a parabolic trajectory, $\frac{x}{x_{0 p}}=\left(\frac{\xi_p}{2}\right)^2$. 

For the entangled Airy photons, the propagation distances of both the signal and the idler photons, $z_s$ and $z_i$, must be taken into account, yielding an effective two-photon propagation governed by $\xi_{s,i}=\frac{z_s+z_i}{k{x_{0}} ^2}$ (see supplementary),

\begin{equation}\label{2}
C(x_s,x_i,\xi_{s,i})\propto \left|Ai\left(\frac{x_s+x_i}{x_0}-\left(\frac{\xi_{s,i}}{2}\right)^2+ia\xi_{s,i} \right)exp\left(a\frac{x_s+x_i}{x_0}-a\frac{\xi_{s,i}^2}{2}\right)\right|^2.
\end{equation}. 

To study the propagation of the entangled Airy photons, we add a quadratic phase with a computer-controlled amplitude onto the SLM (Fig.\ref{fig:1}c), which is equivalent to free space propagation at the far-field\cite{goodman2005introduction} (see supplementary). Remarkably, both the classical Airy beam (Fig.\ref{fig:2}c) and the entangled Airy photons (Fig.\ref{fig:2}d) exhibit negligible diffraction even after a significant propagation distance, $\xi=3$.
The single counts distribution is unaffected by the propagation (Fig.\ref{fig:2}d).

\begin{figure}[H]
\centering
\includegraphics[width=0.35\textwidth]{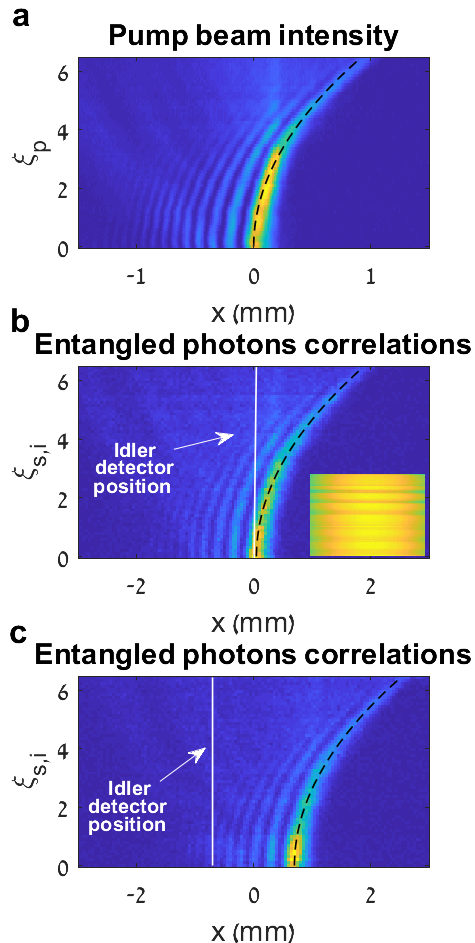}
  \caption{\label{fig:3}  Acceleration of entangled Airy photons. The intensity of the pump beam (a) and the coincidence profile of the entangled Airy photons (b), are measured at different propagation distances emulated by a quadratic phase applied by the SLM. The theoretical parabolic trajectories are marked with a dashed black line, showing good agreement with the experimental results. The single photon distribution remains approximately constant even during propagation (b, inset). In (c), the same measurement is repeated with the stationary single photon detector (a solid white line) displaced by $x=-0.7mm$ along the $x$ axis.}
\end{figure}

Thanks to the correspondence between the classical Airy beam and the entangled Airy photons described by Eq.\ref{2}, the same parabolic trajectory discussed above of the classical Airy beam is expected from the quantum correlations between the photons. The parabolic trajectory of the entangled Airy photons will thus be a function of the two-photon propagation, given by $\frac{x}{x_{0}}=\left(\frac{\xi_{s,i}}{2}\right)^2$ .
Fig.\ref{fig:3}a,b presents the evolution of the correlations between the entangled photons and the intensity of the pump beam during propagation. The dynamics of the entangled Airy photons is strikingly similar to that of the classical Airy beam. The theoretical parabolic trajectory of the main lobe is presented by a black dashed line, showing good agreement with the experimental results.
Finally, we demonstrate the anti-correlated nature of the entangled Airy photons. The structure of the quantum correlations, obtained in Eq.\ref{2}, depends only on the sum of the two photons coordinates, i.e. $C(x_s,x_i,\xi_{s,i})=C(x_s+x_i,\xi_{s,i})$. Therefore, when the stationary heralding detector (idler photon) is shifted along the transverse direction, the Airy distribution that the signal photon collapses to is shifted in the opposite direction. Indeed, we observe that the measured coincidence pattern follows the same parabolic trajectory as before (black dashed line), only with an additional constant shift (Fig.\ref{fig:3}c).

In conclusion, we utilized the unique correspondence between the pump intensity and the entangled photons correlations to generate entangled Airy photons. The spatially entangled Airy photons were shown to have equivalent properties to classical Airy beams, such as free acceleration and negligible diffraction over large propagation distances. In addition, we have shown that the entangled Airy photons posses distinct quantum features, i.e. they are anti-correlated at the far-field and their Airy shape is only revealed in the quantum correlations and not in the single photon distribution. 

We expect spatially entangled Airy photons to benefit future imaging techniques by combining the weak diffraction of Airy beams together with the spatial correlations of quantum illumination \cite{jia2014super,brida2010experimental}. In addition, quantum technologies in which the photons are sent over long propagation distances, such as free-space quantum key distribution (QKD)\cite{liao2017satellite}, could benefit from the reduced diffraction of the Airy photons as well\cite{gu2010scintillation,chu2011evolution}. 
\section*{Funding}
Zuckerman STEM Leadership Program; Israel Science Foundation  (1268/16); United States-Israel Binational Science Foundation (BSF) (2017694).

\printbibliography

@article{berry1979nonspreading,
  title={Nonspreading wave packets},
  author={Berry, Michael V and Balazs, Nandor L},
  journal={American Journal of Physics},
  volume={47},
  number={3},
  pages={264--267},
  year={1979},
  publisher={AAPT}
}

@article{siviloglou2007accelerating,
  title={Accelerating finite energy Airy beams},
  author={Siviloglou, Georgios A and Christodoulides, Demetrios N},
  journal={Optics letters},
  volume={32},
  number={8},
  pages={979--981},
  year={2007},
  publisher={Optical Society of America}
}

@article{siviloglou2007observation,
  title={Observation of accelerating Airy beams},
  author={Siviloglou, GA and Broky, J and Dogariu, Aristide and Christodoulides, DN},
  journal={Physical Review Letters},
  volume={99},
  number={21},
  pages={213901},
  year={2007},
  publisher={APS}
}

@article{ellenbogen2009nonlinear,
  title={Nonlinear generation and manipulation of Airy beams},
  author={Ellenbogen, Tal and Voloch-Bloch, Noa and Ganany-Padowicz, Ayelet and Arie, Ady},
  journal={Nature photonics},
  volume={3},
  number={7},
  pages={395},
  year={2009},
  publisher={Nature Publishing Group}
}

@article{lotti2011stationary,
  title={Stationary nonlinear Airy beams},
  author={Lotti, A and Faccio, D and Couairon, A and Papazoglou, DG and Panagiotopoulos, P and Abdollahpour, D and Tzortzakis, S},
  journal={Physical Review A},
  volume={84},
  number={2},
  pages={021807},
  year={2011},
  publisher={APS}
}

@article{voloch2013generation,
  title={Generation of electron Airy beams},
  author={Voloch-Bloch, Noa and Lereah, Yossi and Lilach, Yigal and Gover, Avraham and Arie, Ady},
  journal={Nature},
  volume={494},
  number={7437},
  pages={331},
  year={2013},
  publisher={Nature Publishing Group}
}

@article{salandrino2010airy,
  title={Airy plasmon: a nondiffracting surface wave},
  author={Salandrino, Alessandro and Christodoulides, Demetrios N},
  journal={Optics letters},
  volume={35},
  number={12},
  pages={2082--2084},
  year={2010},
  publisher={Optical Society of America}
}

@article{minovich2011generation,
  title={Generation and near-field imaging of Airy surface plasmons},
  author={Minovich, Alexander and Klein, Angela E and Janunts, Norik and Pertsch, Thomas and Neshev, Dragomir N and Kivshar, Yuri S},
  journal={Physical review letters},
  volume={107},
  number={11},
  pages={116802},
  year={2011},
  publisher={APS}
}

@article{jia2014super,
  title={Isotropic three-dimensional super-resolution imaging with a self-bending point spread function},
  author={Jia, Shu and Vaughan, Joshua C and Zhuang, Xiaowei},
  journal={Nature photonics},
  volume={8},
  number={4},
  pages={302},
  year={2014},
  publisher={Nature Publishing Group}
}

@article{baumgartl2008micro,
  title={Optically mediated particle clearing using Airy wavepackets},
  author={Baumgartl, J{\"o}rg and Mazilu, Michael and Dholakia, Kishan},
  journal={Nature photonics},
  volume={2},
  number={11},
  pages={675},
  year={2008},
  publisher={Nature Publishing Group}
}

@article{polynkin2009plasma,
  title={Curved plasma channel generation using ultraintense Airy beams},
  author={Polynkin, Pavel and Kolesik, Miroslav and Moloney, Jerome V and Siviloglou, Georgios A and Christodoulides, Demetrios N},
  journal={Science},
  volume={324},
  number={5924},
  pages={229--232},
  year={2009},
  publisher={American Association for the Advancement of Science}
}

@article{maruca2018quantum,
  title={Quantum Airy photons},
  author={Maruca, Stephanie and Kumar, Santosh and Sua, Yong Meng and Chen, Jia-Yang and Shahverdi, Amin and Huang, Yu-Ping},
  journal={Journal of Physics B: Atomic, Molecular and Optical Physics},
  volume={51},
  number={17},
  pages={175501},
  year={2018},
  publisher={IOP Publishing}
}

@article{monken1998transfer,
  title={Transfer of angular spectrum and image formation in spontaneous parametric down-conversion},
  author={Monken, Carlos Henrique and Ribeiro, PH Souto and P{\'a}dua, Sebasti{\~a}o},
  journal={Physical Review A},
  volume={57},
  number={4},
  pages={3123},
  year={1998},
  publisher={APS}
}

@article{walborn2010spatial,
  title={Spatial correlations in parametric down-conversion},
  author={Walborn, Stephen P and Monken, CH and P{\'a}dua, S and Ribeiro, PH Souto},
  journal={Physics Reports},
  volume={495},
  number={4-5},
  pages={87--139},
  year={2010},
  publisher={Elsevier}
}

@article{torres2003preparation,
  title={Preparation of engineered two-photon entangled states for multidimensional quantum information},
  author={Torres, Juan P and Deyanova, Yana and Torner, Lluis and Molina-Terriza, Gabriel},
  journal={Physical Review A},
  volume={67},
  number={5},
  pages={052313},
  year={2003},
  publisher={APS}
}

@article{kovlakov2018oam,
  title={Quantum state engineering with twisted photons via adaptive shaping of the pump beam},
  author={Kovlakov, EV and Straupe, SS and Kulik, SP},
  journal={Physical Review A},
  volume={98},
  number={6},
  pages={060301},
  year={2018},
  publisher={APS}
}

@article{lib2019generation,
  title={Generation of entangled photons with tailored correlations for real-time quantum wavefront shaping},
  author={Lib, Ohad and Hasson, Giora and Bromberg, Yaron},
  journal={arXiv preprint arXiv:1902.06653},
  year={2019}
}

@article{defienne2018coherence,
  title={Spatially-entangled Photon-pairs Generation Using Partial Spatially Coherent Pump Beam},
  author={Defienne, Hugo and Gigan, Sylvain},
  journal={arXiv preprint arXiv:1812.02046},
  year={2018}
}

@article{giese2018influence,
  title={Influence of pump coherence on the quantum properties of spontaneous parametric down-conversion},
  author={Giese, Enno and Fickler, Robert and Zhang, Wuhong and Chen, Lixiang and Boyd, Robert W},
  journal={Physica Scripta},
  volume={93},
  number={8},
  pages={084001},
  year={2018},
  publisher={IOP Publishing}
}

@book{goodman2005introduction,
  title={Introduction to Fourier optics},
  author={Goodman, Joseph W},
  year={2005},
  publisher={Roberts and Company Publishers}
}

@article{greenfield2011accelerating,
  title={Accelerating light beams along arbitrary convex trajectories},
  author={Greenfield, Elad and Segev, Mordechai and Walasik, Wiktor and Raz, Oren},
  journal={Physical Review Letters},
  volume={106},
  number={21},
  pages={213902},
  year={2011},
  publisher={APS}
}

@article{efremidis2019airy,
  title={Airy beams and accelerating waves: an overview of recent advances},
  author={Efremidis, Nikolaos K and Chen, Zhigang and Segev, Mordechai and Christodoulides, Demetrios N},
  journal={Optica},
  volume={6},
  number={5},
  pages={686--701},
  year={2019},
  publisher={Optical Society of America}
}

@article{kovlakov2017bell,
  title={Spatial bell-state generation without transverse mode subspace postselection},
  author={Kovlakov, EV and Bobrov, IB and Straupe, SS and Kulik, SP},
  journal={Physical review letters},
  volume={118},
  number={3},
  pages={030503},
  year={2017},
  publisher={APS}
}

@article{abouraddy2002entangled,
  title={Entangled-photon Fourier optics},
  author={Abouraddy, Ayman F and Saleh, Bahaa EA and Sergienko, Alexander V and Teich, Malvin C},
  journal={JOSA B},
  volume={19},
  number={5},
  pages={1174--1184},
  year={2002},
  publisher={Optical Society of America}
}

@article{broky2008self,
  title={Self-healing properties of optical Airy beams},
  author={Broky, John and Siviloglou, Georgios A and Dogariu, Aristide and Christodoulides, Demetrios N},
  journal={Optics express},
  volume={16},
  number={17},
  pages={12880--12891},
  year={2008},
  publisher={Optical Society of America}
}

@article{brida2010experimental,
  title={Experimental realization of sub-shot-noise quantum imaging},
  author={Brida, Giorgio and Genovese, Marco and Berchera, I Ruo},
  journal={Nature Photonics},
  volume={4},
  number={4},
  pages={227},
  year={2010},
  publisher={Nature Publishing Group}
}

@article{liao2017satellite,
  title={Satellite-to-ground quantum key distribution},
  author={Liao, Sheng-Kai and Cai, Wen-Qi and Liu, Wei-Yue and Zhang, Liang and Li, Yang and Ren, Ji-Gang and Yin, Juan and Shen, Qi and Cao, Yuan and Li, Zheng-Ping and others},
  journal={Nature},
  volume={549},
  number={7670},
  pages={43},
  year={2017},
  publisher={Nature Publishing Group}
}

@article{gu2010scintillation,
  title={Scintillation of Airy beam arrays in atmospheric turbulence},
  author={Gu, Yalong and Gbur, Greg},
  journal={Optics letters},
  volume={35},
  number={20},
  pages={3456--3458},
  year={2010},
  publisher={Optical Society of America}
}

@article{chu2011evolution,
  title={Evolution of an Airy beam in turbulence},
  author={Chu, Xiuxiang},
  journal={Optics letters},
  volume={36},
  number={14},
  pages={2701--2703},
  year={2011},
  publisher={Optical Society of America}
}

\section*{Supplementary information}
\section*{Two-photon propagation}
\def\theequation{S\arabic{equation}}
\setcounter{equation}{0}
 The probability of detecting the signal photon at a position $x_s$ and the idler photon at a position $x_i$ is given by the coincidence pattern, $C(x_s,x_i)=\left|\langle{0}|E_1(x_i)E_2(x_s)|\psi\rangle\right|^2$ \cite{walborn2010spatial}, where $|\psi\rangle$ is the quantum state of the entangled photons, $E_{1,2}$ are the electric field operators and we assume that the pump beam is sufficiently weak so that multi-pair events can be neglected. To study the propagation of the entangled Airy photons, we are interested in the case where the signal and idler photons pass through the 2-f Fourier configuration, and then propagate for distances $z_s$ and $z_i$, respectively. In this case, the electrical field operators in the Heisenberg picture can be written as $E_1(x_i)=\int dqa(q)exp\left(\frac{ik\left(x_i-\frac{f}{k}q\right)^2}{2z_i}\right)$, where f is the focal length of the Fourier lens, k is the wavenumber of the idler photon and $a(q)$ annihilates a photon with transverse momentum $q$\cite{abouraddy2002entangled,walborn2010spatial}. Substituting this expression along with the analogous one for $E_2(x_s)$ into the expression for the coincidence pattern, yields,
 \begin{equation}\label{S1}
C(x_s,x_i,z_s,z_i)\propto \left|\int  dx'W(x')exp\left(-\frac{ik(z_s+z_i)x'^2}{2f^2}\right)exp\left(-i\frac{k}{f}(x_s+x_i)x'\right)\right|^2
\end{equation}
 where $W(x')$ is the profile of the pump beam at the crystal plane. For comparison, the classical pump beam propagating through the same setup will present an intensity pattern of the form $I(x,z)\propto \left|\int  dx'W(x')exp\left(-\frac{ik_pzx'^2}{2f^2}\right)exp\left(-i\frac{k_p}{f}xx'\right)\right|^2$, where $k_p$ is the wavenumber of the pump beam. 
 Thus, by looking at $z_s+z_i$ as an effective two-photon propagation distance, we notice that the result for the entangled photons is equivalent to the classical result, thus leading us to the definition of  $\xi_{s,i}$ given in the main text. One unexpected result from this relation is that moving one of the single photon detectors forwards and the other one backwards by the same amount will not affect the measured correlations.  
 In addition, the expressions obtained here for the coincidence and intensity patterns directly shows that the propagation of the entangled photons and the pump beam are equivalent to the addition of quadratic phase with a coefficient $-\frac{k(z_s+z_i)}{2f}$ and $-\frac{k_pz}{2f}$ in the crystal plane, respectively. We can therefore emulate propagation by utilizing the SLM for applying quadratic phase with a variable curvature.

\end{document}